\documentclass[aps, prb, reprint, twocolumn, showkeys, showpacs,floatfix]{revtex4-2}
\usepackage{graphicx}
\usepackage{bm, amsmath, amssymb}
\usepackage{xcolor}
\usepackage[normalem]{ulem} 

\begin{document}

\title[\today]{Fast and Slow Sound Excitations in Nematic Aerogel in
superfluid $^{3}$He}
\author{A.M. Bratkovsky}
\affiliation{Kapitza Institute for Physical Problems, Moscow 119334, Russia}

\begin{abstract}
Nematic aerogel (nAG) supports so-called polar phase in liquid $^{3}$He. The
experiments  [Dmitriev \textit{et al.}, JETP Lett. \textbf{112}, 780
(2020)] showed that the onset of polar phase inside the nAG is accompanied
by emergence of a sound wave with frequency quickly growing with cooling
down from transition temperature and reaching a plateau. To describe this
behavior, we start by calculating the elastic properties of the dry nematic
AG that appear to depend only on Young's modulus of the parent material
(e.g. mullite), the volume fraction of the solid phase $\psi~(=5\%)$ and the
aspect ratio of the representative volume of nAG. The elastic constants are
then used to solve elasto-hydrodynamic equations for various sound
vibrations of nAG filled with $^{3}$He. The (isotropic) first sound and
anisotropic second sound in the polar phase are strongly hybridized with
fourth sound and standard elastic modes in nAG. The hybrid second and the
transverse fourth sound start with zero velocity at the transition, similar
to pure $^{3}$He, and quickly grow with lowering temperature until they hit
the sample finite size cutoff.
\end{abstract}

\date{\today }
\email{alex.bratkovski@gmail.com}
\startpage{1}
\endpage{1002}
\maketitle


\section{Introduction}

Helium three, $^{3}$He, is the only topological superfluid accessible in
laboratory conditions. It is intrinsically anisotropic sharing some features
with liquid crystals\cite{leggett75}. The polar phase of superfluid $^{3}$%
He, that was expected for quite a while to be stabilized in anisotropic
aerogel\cite{ikeda06}, was found and studied extensively since 2015%
\cite{dmit15,dmit16,dmit20}. Very strong anisotropy of
filamentary porous media housing $^{3}$He is required, and it was provided
by new type of nematic aerogel (nAG). The signatures of the polar phase have
been found in experiments where nAG\ was attached to a \emph{vibrating wire}
(VW.) This technique has been used extensively to study $^{3}$He, see review 
\cite{dmitufn24}. In addition to finding the polar phase, Dmitriev {\em  et al.}
have found the so-called \emph{beta-phase} of $^{3}$He in magnetic field~\cite%
{dmitBeta21}. Both polar- and beta-phases are equal spin pairing (ESP) so
that the condensate of the Cooper pairs is made of aligned spin pairs like $%
\left\vert \uparrow \uparrow \right\rangle $. Wave function of the pair in
momentum $k-$space is the $2 \times 2$ bispinor $\Psi _{\sigma \sigma ^{\prime }}( 
\boldsymbol{\hat{k}}) $ ($\boldsymbol{\hat{k}}$ is the unit $k-$%
vector), $\sigma ,\sigma ^{\prime }$ the spin quantum numbers of $^{3}$He
atoms in the pair). One may construct a vector $d_{\lambda }(\boldsymbol{%
\hat{k}})=-i~$tr$\left( \Psi _{\sigma \sigma ^{\prime }}\sigma _{2}\sigma
_{\lambda }\right) \equiv A_{\lambda j}\hat{k}_{j}$ , with $d_{\lambda }$
living in spin space, $\sigma _{\lambda }=\left( \sigma _{1},\sigma
_{2},\sigma _{3}\right) $ the vector of Pauli matrices. Its $k-$dependence
must be linear for p-wave orbital state of the pair, $L=1$, and this explains
the last equality. The matrix $A_{\lambda j}$ (the order parameter) defines
the properties of particular phases that are formed at a certain pressure,
boundary conditions, and internal perturbations like interactions with
embedded aerogel. The so-called B-phase has the simplest structure, $%
d_{\lambda }(\boldsymbol{\hat{k}})\propto \boldsymbol{\hat{k}}$, with
isotropic gap in spectrum of Bogoliubov quasiparticles, $\Delta _{%
\boldsymbol{k}}\propto \left\vert d_{\lambda }(\boldsymbol{\hat{k}}%
)\right\vert ^{2}=\mathrm{const}$. The $^{3}$He$\ $interactions with
embedded \emph{nematic} aerogel favor 'uniaxial' $d_{\lambda }(\boldsymbol{%
\hat{k}})\propto \hat{k}_{z}=\cos \theta $, where $\theta $ the polar angle
with the \emph{director} of the nematic aerogel $\boldsymbol{m}$\ that we
select as the $z-$axis. This yields the gap $\Delta _{k}\propto \hat{k}%
_{z}^{2}=\cos ^{2}\theta $ with the Dirac nodal line along the equator of
the fermi surface, $\theta =\pi /2$. The spin \emph{projection} of the
Cooper pair along z-axis vanishes, $S_{z}=0$ (total spin of the pair is $S=1$%
) since its wave function is $\Psi _{\text{pair}}^{\text{polar}}\propto ~%
\hat{k}_{z}(\left\vert \uparrow \uparrow \right\rangle +\left\vert
\downarrow \downarrow \right\rangle )$ for $\boldsymbol{d}\propto
(0,-ik_{z},0)$ using z-axis quantization. In variance with the \emph{polar}
phase, its sibling the \emph{beta} phase only comprises a pair of up-spins \ 
$\left\vert \uparrow \uparrow \right\rangle $ and no down-spins in external
magnetic field, with $\boldsymbol{d}\propto (-k_{z},-ik_{z},0),$\ thus
making it similar to $A_{1}$ phase. The $\beta$ phase also has a gap with a nodal line 
running along equator of fermi surface.  The strongly anisotropic nematic AG\
orients the Cooper pairs so that the superfluid phase has maximal value of
the gap at the $\pm \hat{z}$ poles, where $\hat{z}$ is the unit vector
pointing along the growth direction (parallel to the director $\boldsymbol{m}
$).

Both polar- and beta-phases of $^{3}$He have been discovered using
approximately (3 mm)$^{3}$ nematic AG glued to an apex of an arch-shaped
vibrating wire (VW)\ subject to the driving AC\ current in external magnetic
field. The sample oscillated along the strands. The strands have diameters
about $14$ nm and average separation between them about $60$ nm (the volume
fraction of the solid part $\psi\approx5\%$.) The observed frequency range has been $%
f=500-1600$ Hz. Importantly, the viscous penetration depth $\delta\approx0.3$
mm was much larger than the
spacing between the strands even near $T_{c}$ at $P=29.3$ bar \cite{dmit20} 
meaning that the \emph{normal component} in the
above experiment is always \emph{clamped to the aerogel}. Since nAG\ drags
along the normal part, the overall dynamics of vibrations changes
drastically especially because the density of superfluid motion parallel to
the strands is at least three times that perpendicular to the strands, $%
\rho_{\parallel}^{s}/\rho_{\perp}^{s}\geq3,$ near critical temperature of
interest to us (the lower limit corresponds to a weak coupling
approximation.) Thus, the effective mass density involved in various sound
vibrations of the combined system $^{3}$He-nAG depends on the polarization
of those vibrations. This should be contrasted with $^{4}$He \cite{mckenna91}
and $^{3}$He \cite{g99parp05} in \emph{isotropic} silica aerogel.

In experiments by Dmitriev {\em et al.} \cite{dmit20}, 
	the vibrating wire with attached nAG has
exhibited the main resonance in the normal phase above $T_{c}$. Then
a rapid decrease of the width has been observed indicating a superfluid
transition in bulk $^3$He at $T=T_c$. On further cooling,
the second resonance has appeared due to the superfluid transition of 
$^{3}$He into the polar phase in the oscillating sample at $T_{ca}\approx
0.989T_{c}$. Although the authors have not been able to observe a
clear resonance peak at frequencies lower than 470 Hz, they assumed that on
cooling from $T=T_{ca}$ the frequency of the second mode
rapidly grows from 0 to about 1600 Hz \cite{dmit20}.

As far as the origin of observed \emph{additional resonance}, the
authors have speculated that the 'slow mode' should correspond to some soft
deformations of nAG perpendicular to strands since the sample is very stiff
along the strands. This is correct and, in fact, \emph{any} sound but the 
\emph{longitudinal} fourth sound is slow, as we calculate below without any
fitting parameters for all vibrational modes of combined $^{3}$He-nAG
system. In fact, the nAG\ itself supports plenty of slow modes being highly
anisotropic extremely porous material with tiny solid state volume fraction $%
\psi=5\%.$ Without solving for the elastic properties of nematic aerogel
first, one is left with large number of unknown fitting parameters to
describe the dynamics of anisotropic superfluid encased in highly
anisotropic medium, see prior works \cite{brand20n22,surfA}. Below, we shall
calculate all the elastic constants of nAG first and then solve the combined
elasto-hydrodynamic equations of motion for all hybrid vibrational modes and
give results for the polar phase for sound waves propagating along and
perpendicular to the strands. We shall focus on properties of uniform $^{3}$%
He-nAG and discuss how the finite size of a nAG sample affects the results.

We shall see below that the sample supports many \emph{slow modes} due to
elastic vibrations of the AG skeleton that drags along the highly
anisotropic normal part of the polar $^{3}$He. They exist above and below
transition and weakly depend on temperature with velocities $U\sim1-10$ m/s.
Below transition, there appear the \emph{second and fourth sound modes }%
hybridized with aerogel skeleton vibrations. The hybrid second sounds start
with \emph{zero velocity,} $U_{2a}(T_{ca})=0$ that increases rapidly with
lowering $T$. The \emph{hybrid fourth sound} has zero velocity at $T_{ca}$ 
\emph{only} for transversal modes propagating \emph{perpendicular} to the
strands, while the hybrid longitudinal fourth sound would have finite and
large velocity on the order of $u_{1}\gtrsim100$ m/s at all temperatures,
including $T_{ca}$. This mode is too fast to be excited in small aerogel
sample with linear size $L\ll\lambda/2$ (half wavelength of the sound wave)
and, therefore, is not responsible for the observed resonances \cite{dmit15,
dmitufn24}. All modes are described by simple quadratic equations with
parameters depending on the propagation direction and polarization of the
waves reflecting the anisotropy of both the polar phase of $^{3}$He and nAG
defining the anisotropic hybrid second and fourth sounds.  The above simple
scenario for 'slow mode' resonance is different from the interesting one
studied in Ref.\cite{surfA} that considered volume-conserving shape
oscillations of finite AG sample coupled to a soft mode related to chemical
potential coupling to an axial strain. In the present model, the
oscillations of the chemical potential are accounted for in a usual way
through oscillations of pressure and temperature (two leading terms) in the
sound waves\cite{LL6} and the sound velocities are given as solutions of
simple quadratic equation with no singular denominators.

\section{Elastic properties of nematic aerogel}

The nematic AG consists of rigid polycrystalline strands of mullite or other
inorganic material like Al$_{2}$O$_{3}$ or Al$_{2}$O$_{3}$-ZrO$_{2}$. It is
a high-porosity anisotropic network structure that is stiff along the growth 
$z-$direction and is elastically isotropic in plane perpendicular to the
strands. It would then be characterized by five elastic constants like any
transversely isotropic system. The nematic aerogel contains rather straight
strands with small waviness along the growth direction, Fig.1(b). They touch
along z-axis to form an elastic skeleton with typical separation between the
rigid joints $c$ that is much larger than the typical spacing between the
strands $a\sim 60$ nm, i.e. $c/a\gg 1.$ The typical diameter of the strands
is $2r\leq 14$ nm. Our estimates below show that the effective elastic
constants of nAG depend only on the Young's modulus of the strands' material 
$E$, the aspect ratio $c/a,$ and the solid fraction $\psi $ (about 5\% in
mullite nAG\cite{dmit20}.) Thus, it appears that the present expressions for
the elastic constants of nematic AG are rather general. This is in the same
vein as the elastic properties of highly porous cellular materials that
follow simple scaling laws\cite{gibcell}.

\begin{figure}[ptb]
\centering\includegraphics[width=0.4\textwidth]{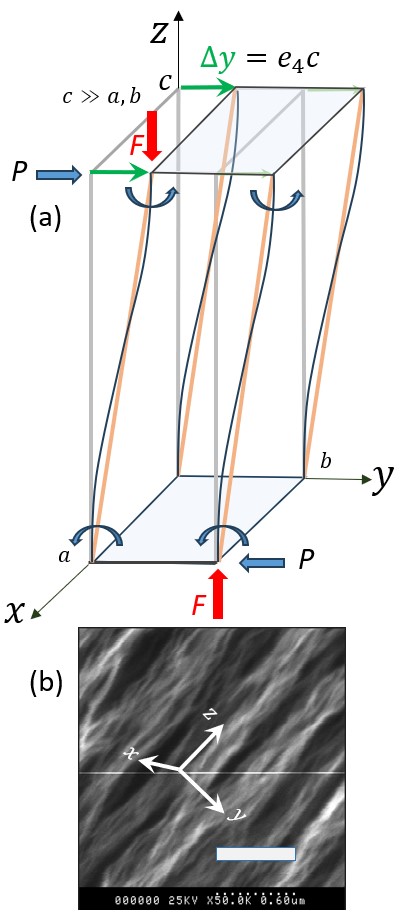} 
\caption{(Color online)
(a) Representative volume of nematic aerogel used to estimate its elastic constants
viewed as an equivalent orthorhombic cell with fused nodes.
The strands are fused together over the typical distance $c \gg a,b$ , where the parameters $a$ and $b$ are
typical separation between the strands ($\sim$60 nm in Ref.~\cite{dmit20}.)
$P$ and $F$  mark the shear and the axial forces resulting in shear strain $e_{23}\equiv e_5$
and axial strain (not shown). Under the load, the fused strands bend and rotate about the nodes.
The resulting bending torques on the strands (elastic Euler-Bernoulli "beams") are shown by curved arrows.
(b) The STEM picture of nematic aerogel (courtesy V.V. Dmitriev, A.A.Soldatov, and A.N. Yudin.)
The size bar is 0.6 $\mu$m.
}\label{Fig:ucell_w_TEM}
\end{figure}

To find the effective elastic constants, consider the simplest
representative periodic orthorhombic unit cell of AG viewed as an elastic 
\emph{frame} in Fig.~1 with sides $a\approx b\ll c.$ We assume that the
strands are rigidly joined at the nodes in accord with TEM data  ensuring bending moment continuity.
The periodic model is then subjected to uniform strain with strain tensor $%
e_{ij}=\{e_{xx},e_{yy},e_{zz},e_{yz},e_{xz},e_{xy}\}$ or, 
equivalently, $\{e_{1},e_{2},e_{3},e_{4},e_{5},e_{6}\}$ in Voigt notations.
The strands will exhibit axial and shear (bending) deformations. One
calculates the total energy of the deformed unit cell and equates it to the
elastic energy $W_{el}=\frac{1}{2}\upsilon_{c}C_{ijlm}e_{ij}e_{lm}$ [J]
through the elastic constants tensor $C_{ijlm}$ [J/m$^{3}$] and determines
all its components in the linear elastic approximation ($\upsilon_{c}$ [m$%
^{3}$] is the unit cell volume). We assume Einstein summation rule over
repeated indices. This will suffice to study the sound excitations that we
are focusing on. Other e.g. \emph{flexural} modes of the strands with $%
\omega\propto k^{2}$ dispersion could be accounted for when needed.

For axial strain along $z-$direction involving $c-$strands (Fig.1), the
latter will deform by $\delta=e_{3}c$ with corresponding energy 
\begin{equation}
W_{axial}(e_{zz})=\frac{1}{2}EAc\left( \frac{\delta}{c}\right) ^{2}=\frac {1%
}{2}EAc~e_{3}^{2},
\end{equation}
where $E$ is the Young's modulus of the strand material ($E=150$ GPa for
mullite) and $A=\pi r^{2}$ the cross sectional area of the strand. Analogous
results hold for strains along $a-$ and $b-$strands.

The strains $e_{4\div6}$ result in bending of the strands and rigid
rotations of the nodes (Fig.1). Consider $e_{4}\equiv e_{yz}:$ the top of
the unit cell would shift by $\Delta y=e_{yz}c$ $\equiv e_4c$.
The force resulting in this deformation is $P=12EI~\Delta y/c^{3}$ , $I=\pi
r^{4}/4$ \cite{LL7}. The corresponding energy is 
\begin{equation}
W_{bend}(e_{yz})=\frac{1}{2}P~\Delta y=6EI\frac{e_{4}^{2}}{c}.
\end{equation}

The total elastic energy%
\begin{align}
W_{tot} & =\frac{1}{2}\upsilon_{c}c_{\alpha\beta}e_{\alpha}e_{\beta }  \notag
\\
& =\frac{1}{2}EA\left( ae_{1}^{2}+be_{2}^{2}+ce_{3}^{2}\right)  \notag \\
& +6EI\left[ \left( \frac{1}{b}+\frac{1}{c}\right) e_{4}^{2}+\left( \frac{1}{%
a}+\frac{1}{c}\right) e_{5}^{2}+\left( \frac{1}{a}+\frac{1}{b}\right)
e_{6}^{2}\right] ,
\end{align}
where $\upsilon_{c}=abc$ is the unit cell volume, $\alpha,\beta=1\div6$ are
the Voigt indices. This yields the elastic constants%
\begin{align}
c_{11}(c_{22},c_{33}) & =EAa(b,c)/\upsilon_{c},  \notag \\
c_{44} & =6EI\left( \frac{1}{b}+\frac{1}{c}\right) /\upsilon_{c},  \notag \\
c_{55} & =6EI\left( \frac{1}{a}+\frac{1}{c}\right) /\upsilon_{c},  \notag \\
c_{66} & =6EI\left( \frac{1}{a}+\frac{1}{b}\right) /\upsilon_{c},
\label{eq:All_c_ab}
\end{align}
with other ones being negligible, meaning that the Poisson coefficients of
the aerogel network structure are close to zero. Using the small parameter $%
a/c\approx b/c\ll1,$ one obtains the important expressions for the 'axial'
elastic constants,%
\begin{align}
c_{33} & \approx E\psi,  \notag \\
c_{11} & \approx c_{22}\approx E\psi\frac{a}{c}\ll c_{33},
\end{align}
and for the shear constants,%
\begin{align}
c_{44} & \approx\frac{3}{2\pi}E\psi^{2}\frac{a}{c},~c_{55}\approx
c_{44},~c_{66}\sim2c_{44},  \label{eq:c44} \\
c_{44,55,66} & \ll c_{11,22}\ll c_{33}.  \label{eq:44<<11<<33}
\end{align}
The shear constants contain an extra factor $\psi,$ the small volume
fraction of the solid material ($\sim5\%$), and this is reflected in very 
\emph{soft 'shear' or 'bending' modes} for sound propagating in nematic
aerogel. Since the Poisson coefficients for the network structure are
negligible ($c_{12}\ll c_{44})$, the matrix of Voigt elastic constants is
diagonal to a good approximation, 
\begin{align}
& c_{\alpha\beta}=\mathrm{diag}\left(
c_{11},c_{22},c_{33},c_{44},c_{55},c_{66}\right)  \notag \\
& \approx\mathrm{diag}\left( E\psi\frac{a}{c},E\psi\frac{a}{c},E\psi ,\frac{3%
}{2\pi}E\psi^{2}\frac{a}{c},\frac{3}{2\pi}E\psi^{2}\frac{a}{c},\frac{3}{\pi}%
E\psi^{2}\frac{a}{c}\right) .
\end{align}
In the last line we shall omit the numerical factors on order unity.

\section{Vibrational modes in (polar) $^{3}$He-nAG}

The polar phase is a non-chiral strongly anisotropic superfluid, as follows
from its condensate wave function discussed above. It has the maximal
superfluid gap oriented along the strands while it vanishes perpendicular to
the strands. The densities of both superfluid and normal motions are
tensorial quantities assembled into the total density $\rho$ of $^{3}$He,%
\begin{equation}
\rho\delta_{ij}=\rho_{ij}^{n}+\rho_{ij}^{s},  \label{eq:rhodij}
\end{equation}
where $\rho_{ij}^{n(s)}=\rho_{\parallel}^{n(s)}\hat{z}_{i}\hat{z}%
_{j}+\rho_{\parallel}^{n(s)}\delta_{ij}^{\perp},$ \ $\delta_{ij}^{\perp}=%
\delta_{ij}-\hat{z}_{i}\hat{z}_{j}$.\ This structure is strongly reflected
in the sound velocities. Near transition temperature (Ginzburg-Landau
regime)\ the anisotropy is large, $\rho_{\parallel}^{s}/\rho_{\perp}^{s}%
\geq3.$ This simply reflects the presence of a nodal line on the gap, the
equator of the fermi surface in the plane perpendicular to the strands.
There, the quasiparticle excitations (the 'normal' part of the fluid)\ are
easily excited thus suppressing the density of condensate particles near
equator. Note that as $T\rightarrow0,$ the excitations would die out i.e.
the 'normal' part of the fluid would vanish $\rho_{ij}^{n}\propto T^{2}$ and
the condensate would become \emph{isotropic} and the 'superfluid' part of
density would become equal to the total density. Indeed, $%
\rho_{ij}^{s}(T=0)=\rho\delta _{ij}$, Eq.~(\ref{eq:rhodij}), in spite of the
gap being anisotropic.

We shall use the set of conservation laws for two-fluid hydrodynamics\cite%
{LL6} with density conservation for aerogel $\rho_{a},$ $^{3}$He $\rho,$ the
entropy per unit mass $s$ carried by the normal motion, and the momentum
density $j_{i}$ per unit mass of the $^{3}$He-nAG combined system with
density tensor for the normal motion $\rho_{a}\delta_{ij}+\rho_{ij}^{n},$
and $\rho_{ij}^{s}$ for the superfluid motion\cite%
{mckenna91,brand20n22,surfA},%
\begin{align}
\partial_{t}\rho_{a}+\operatorname{div}\left( \rho_{a}\boldsymbol{v}^{n}\right) & =0,%
\partial_{t}\rho+\operatorname{div}\boldsymbol{g} & =0, \\
\partial_{t}\left( \rho s\right) +\operatorname{div}\left( \rho s\boldsymbol{v}%
^{n}\right) & =0,
\end{align}
where $\boldsymbol{g}$ is the momentum density of $^{3}$He, $g_{i}=\rho
_{ij}^{n}v_{j}^{n}+\rho_{ij}^{s}v_{j}^{s},$and $v_{j}^{n(s)}$ the velocity
of the normal (superfluid) motion. The superfluid motion is irrotational, $%
\operatorname{curl}v^{s}=0,$ and its equation of motion is 
\begin{align}
\dot{v}_{j}^{s} & =-\nabla_{j}\tilde{\mu}, \\
\tilde{\mu} & =-s\tilde{T}+\frac{1}{\rho}\tilde{p}+...  \label{eq:mu_T_p}
\end{align}
where $\tilde{\mu},$ $\tilde{p}$, $\tilde{T}$ are the variations of the
chemical potential per unit mass, pressure, and temperature in the wave
versus equilibrium, $\rho$ the density of $^{3}$He in equilibrium.
Ellipses mark other possible terms allowed by symmetry \cite{brand20n22},%
\cite{surfA} that we shall ignore since they are\ supposed to be small in
comparison with the leading terms in (\ref{eq:mu_T_p}). The momentum density
conservation (second Newton's law) reads%
\begin{align}
\partial_{t}j_{i}+\nabla_{j}\Pi_{ij} & =0, \\
j_{i} &
=\rho_{a}v_{i}^{n}+g_{i}=\rho_{a}v_{i}^{n}+\rho_{ij}^{n}v_{j}^{n}+%
\rho_{ij}^{s}v_{j}^{s} \\
\Pi_{ij} & =p\delta_{ij}-\sigma_{ij}^{R},\text{ }
\end{align}
where the elastic reaction of nAG upon $^{3}$He\ is accounted for by the
elastic stress tensor $\sigma_{ij}^{R}=C_{ijlm}\nabla_{l}u_{m}$ where $u_{m}$
is the strain of the nAG skeleton to which the normal component of $^{3}$He
is clamped and $C_{ijlm}$ the tensor of elastic constants for nAG. For
harmonic motion, when velocities $v^{n(s)}$, and the variations of densities 
$\tilde{\rho},$ $\tilde{\rho}_{a}$, entropy $\tilde{s},$\ pressure $\tilde{p}
$, and temperature $\tilde{T}$ in the wave are all proportional to $\exp(i%
\boldsymbol{kr}-i\omega t)$ specified as $\exp ik(z-Ut)$ for a wave
propagating in e.g. $z-$direction with velocity $U,$\ $%
\sigma_{ij}^{R}=C_{ijlm}\nabla_{l}\left( \frac{v_{m}^{n}}{-i\omega}\right) $.

To find sound velocities, one is solving the above system linearized with
respect to $v_{j}^{n(s)},$ $\tilde{\rho},$ $\tilde{\rho}_{a}$, $\tilde{p}$,
and $\tilde{T}$ \cite{LL6}. The parameters below without tilde are those for
equilibrium: 
\begin{align}
& \partial_{t}\tilde{\rho}+\rho_{ij}^{n}\nabla_{i}v_{j}^{n}+\rho_{ij}^{s}%
\nabla_{i}v_{j}^{s}=0, \\
& s\partial_{t}\tilde{\rho}+\rho\partial_{t}\tilde{s}+\rho s\operatorname{div}%
\boldsymbol{v}^{n}=0, \\
& \dot{v}_{j}^{s}=s\nabla_{j}\tilde{T}-\frac{1}{\rho}\nabla_{j}\tilde{p}, \\
& \rho_{a}\dot{v}_{i}^{n}+\rho_{ij}^{n}\dot{v}_{j}^{n}+\rho_{ij}^{s}\dot {v}%
_{j}^{s}+\nabla_{i}\tilde{p}-\nabla_{j}\left[ C_{ijlm}\nabla_{l}\left( \frac{%
v_{m}^{n}}{-i\omega}\right) \right] =0,  \label{eq:_jmom_dot}
\end{align}
where dot marks the time derivative. A note is in order with regards to Eq.~(%
\ref{eq:_jmom_dot}): it correctly recovers the limiting cases of sound in a 
\emph{dry aerogel} where $\rho_{ij}^{n}=\rho_{ij}^{s}=0,$as well as first
and (anisotropic) second sound in pure $^{3}$He. Thus, the dry nAG shows
three sound branches, one longitudinal and two transversal, the latter being
degenerate for propagation direction along the growth axis, as expected\cite%
{LL7}. Below, we shall drop the effects of (negligible) thermal expansion: 
\begin{align}
\tilde{\rho} & =\left( \frac{\partial\rho}{\partial p}\right) _{T}\tilde{p}%
+\left( \frac{\partial\rho}{\partial T}\right) _{p}\tilde{T}\approx\left( 
\frac{\partial\rho}{\partial p}\right) _{s}\tilde{p}=\frac{\tilde{p}}{%
u_{1}^{2}},  \label{eq:rhoTilde} \\
u_{1}^{2} & \equiv\left( \frac{\partial p}{\partial\rho}\right) _{s},
\label{eq:u1_1st_sound} \\
\tilde{s} & =\left( \frac{\partial s}{\partial p}\right) _{T}\tilde {p}%
+\left( \frac{\partial s}{\partial T}\right) _{p}\tilde{T}\approx \frac{c_{p}%
}{T}\tilde{T},  \label{eq:sTilde}
\end{align}
where $u_{1}$ is the isotropic velocity of first (pressure) sound, $c_{p}$
the specific heat of $^{3}$He.

\subsection{Sound propagating along the strands with $\hat{k}=(001)$}

\begin{figure}[ptb]
\centering
\includegraphics[width=0.3\textwidth]{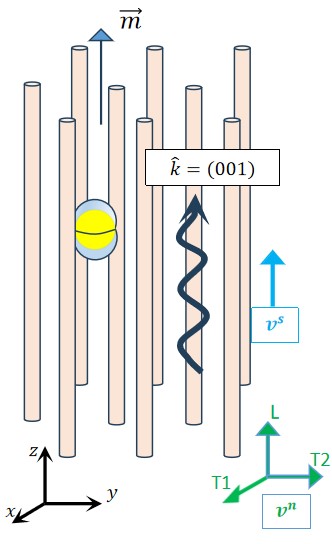} 
\caption{(Color online)
Schematic of sound wave propagating along the strands of nematic aerogel filled with the polar phase  of $^3$He that shows the Dirac nodal line at the equator plane perpendicular to the nematic director $\vec{m}$. The irrotational superfluid motion with velocity $\textbf{v}^s$ follows  the wave  vector $\hat{k}=(0,0,1)$. The normal velocity has one longitudinal and two transversal possible orientations ($L, T1, T2$), The modes $T1$ and $T2$ are degenerate since the nematic aerogel is the transversely isotropic system.
}\label{Fig:Lwave}
\end{figure}

We start by looking for a longitudinal (L) wave $\boldsymbol{v}%
_{L}^{n}\propto\hat{z}\exp ik(z-Ut)$ and two degenerate transversal waves
(T) \ $\boldsymbol{v}_{T1}^{n}\propto\hat{x}\exp ik(z-Ut),$ $\boldsymbol{v}%
_{T2}^{n}\propto\hat{y}\exp ik(z-Ut),$ where the hats mark unit vectors in a
particular direction (choice of x- and y-axes in plane perpendicular to the
strands $\hat{z}$ is arbitrary), Fig.~\ref{Fig:Lwave}.

\subsubsection{L-wave along the strands}

The system of linearized equations reads%
\begin{align}
& -U\tilde{\rho}+\rho _{\parallel }^{n}v^{n}+\rho _{\parallel }^{s}v^{s}=0,
\label{eq:rhoDotLin} \\
& sU\tilde{\rho}+\rho U\tilde{s}-\rho sv^{n}=0,  \label{eq:(rho.s)DotLin} \\
& Uv^{s}+s\tilde{T}-\frac{1}{\rho }\tilde{p}=0,  \label{eq:vsDotLin} \\
& \left( \rho _{a}+\rho _{\parallel }^{n}\right) Uv^{n}+\rho _{\parallel
}^{s}Uv^{s}-\tilde{p}-c_{33}\frac{v^{n}}{U}=0.  \label{eq:EulerLin}
\end{align}%
Again, the '\emph{longitudinal} densities' $\rho _{\parallel }^{n(s)}$ are
involved in dispersion of the waves propagating \emph{along} the strands. Its
solution also yields $\tilde{\rho}_{a}$ through $U\tilde{\rho}_{a}=\rho
_{a}v^{n}$. Note that for the longitudinal wave propagating \emph{along the
strands} only the densities of \ $^{3}$He \emph{parallel to the strands} get
involved. Using Eqs.(\ref{eq:rhoTilde}), (\ref{eq:sTilde}), this system
reduces to a quadratic dispersion equation for the sound velocity $U^{2}$,

\begin{equation}
\left( U^{2}-u_{1}^{2}\right) \left( U^{2}-u_{2\parallel }^{2}\right) +\frac{%
\rho _{a}}{\rho _{\parallel }^{n}}\left( U^{2}-u_{33}^{2}\right) \left(
U^{2}-u_{4\parallel }^{2}\right) =0,  \label{eq:U2_Lz}
\end{equation}%
Here,
\begin{align}
u_{2\parallel }^{2}& \equiv \frac{Ts^{2}}{c_{p}\rho _{\parallel }^{n}}\rho
_{\parallel }^{s}, \\
u_{4\parallel }^{2}& \equiv u_{1}^{2}\frac{\rho _{\parallel }^{s}}{\rho }+%
\frac{\rho _{\parallel }^{n}}{\rho }u_{2}^{2},
\end{align}%
are the (generally anisotropic) second and fourth sound velocities. Above,
we have defined $\rho _{a}u_{33}^{2}\equiv c_{33},$ $c_{33}=E\psi =$\ $150$
GPa$~\times 0.05=7.5$ GPa, $\rho _{a}=\rho _{M}\psi $ where $\rho
_{M}=2.80-3.1$ g/cc is the density of the bulk (ceramic) mullite, giving the 
\emph{characteristic} sound velocity in dry nGA $u_{33}\simeq \sqrt{E\psi
/\left( \rho _{M}\psi \right) }=220$\ m/s (the solid fraction $\psi $ drops
out.) The wave with such large velocity could not be excited in a small nAG\
sample with linear size about $L=3$~mm. Note that this aerogel mode exists above $T_{ca}$
but is too fast to be excited in 3~mm sample (see below.)

The dispersion equation (\ref{eq:U2_Lz}) differs from the one for $^{4}$He
in isotropic aerogel\cite{mckenna91} by account for anisotropy of both the
polar phase and the nematic aerogel. If the aerogel is absent, $\rho _{a}=0,$
the second term in Eq.(\ref{eq:U2_Lz}) vanishes and we recover
the Landau equation for the first and (anisotropic in $^{3}$He) second
sound, $\left( U^{2}-u_{1}^{2}\right) \left( U^{2}-u_{2\parallel
}^{2}\right) =0$ \cite{LL6}. In another limiting case of aerogel mass density $\rho
_{a}\gg \rho _{\parallel }^{n},$ the second term in (\ref{eq:U2_Lz})
dominates, and one recovers the fourth sound in aerogel that acts as a
superleak, yet elastic not the rigid one, plus the pressure sound in dry
aerogel. The latter is high speed and is hardy relevant for the observations%
\cite{dmitufn24}.

A short note on \emph{size effect
cutoff}: a sound wave with velocity $U$ could be excited at frequency $f$ in
a sample with size $L$ provided that $U/(2f)\leq L$ or $U\lesssim U_{c}\sim
10$ m/s for frequencies $f\lesssim 1600$ Hz \cite{dmitufn24}. We shall see
below that there are (i) slow modes supported by elastic reaction of the
aerogel skeleton that exist above and below transition into the polar phase
and (ii) the modes involving the condensate that emerge at $T_{ca}$ with
zero velocity that then rises very quickly to the cutoff value $U_{c}$, 
Fig.4. This is likely to be signaled by observed plateau in
mechanical resonance of the vibrating wire \cite{dmit20,dmitufn24}
(see below.)

The general solutions to the quadratic dispersion law (\ref{eq:U2_Lz}) for
the hybrid second sound $U_{2a}^{L}$ and the hybrid fourth sound $U_{4a}^{L}$%
\ are readily found. The remarkable result is that the \emph{hybrid second
sound} $U_{2a}^{L}$ still vanishes at critical temperature $T_{ca}$ of
superfluid transition in aerogel in spite of persistent elastic force
exerted by aerogel. Near $T=T_{ca},$%
\begin{align}
\left( U_{2a}^{L}\right) ^{2}& \simeq \tau u_{1}^{2}\frac{\rho
_{a}u_{33}^{2}+\rho _{\parallel }^{n}u_{20}^{2}}{\rho _{a}u_{33}^{2}+\rho
_{\parallel }^{n}u_{1}^{2}}  \notag \\
& \approx \tau u_{1}^{2}\frac{\rho _{a}u_{33}^{2}}{\rho _{a}u_{33}^{2}+\rho
u_{1}^{2}},  \label{eq:U2a_nearTca} \\
\left( U_{4a}^{L}\right) ^{2}& \simeq \frac{\rho _{n}u_{1}^{2}+\rho
_{a}u_{33}^{2}}{\rho _{a}+\rho }+C\tau ,  \label{eq:U4a_nearTca}
\end{align}%
where $\tau \equiv 1-T/T_{ca}\ll 1$, $\rho _{\parallel }^{n}\approx \rho $
(density of $^{3}$He) near $T_{ca}$, $u_{20}$ the second sound velocity at
low temperatures, and $C$ the constant. In the above estimate we accounted
for $u_{1}\ggg u_{20}$ ($u_{1}\sim 300$ m/s\cite{surfA}, while $%
u_{20}\lesssim 3$ cm/s\cite{kojima85}) . Note the important \emph{large} $u_{1}$ \emph{factor}
in $U_{2a}$ and that it  emerges at $T=T_{ca}$, Eq.~(\ref%
{eq:U2a_nearTca}), as $U_{2a}\simeq u_{1}\sqrt{\tau }.$ This likely relates
to the observations where vibrating wire resonant frequency sharply
rises upon cooling away from $T_{ca}$ and quickly hits a plateau as a
function of temperature caused by the size effect cutoff at $U_{2a}\left(
T \right) \simeq U_{\text{cutoff}}$. The hybrid fourth sound velocity $%
U_{a4}^{L}$ remains finite and large, Eq.~(\ref{eq:U4a_nearTca}), 
see Fig.~\ref{Fig:schema}. It
cannot be excited in the small mm-size aerogel nAG\ sample.

\subsubsection{Transversal waves along the strands, $\boldsymbol{\hat{k}}%
=(001),\boldsymbol{v}^{n}\perp\hat{z},$ $\boldsymbol{v}^{s}\parallel\hat{z}$}

In this case the normal velocity $\boldsymbol{v}^{n}\propto \{\hat{x},\hat{y}%
\}$ $\exp ik(z-Ut),$ normal to the polarization of superfluid velocity that
is always pointing along the $k-$vector, $\boldsymbol{v}^{s}\propto \hat{z}%
\exp ik(z-Ut).$ Compared to the case of sound propagating along the strands (%
\ref{eq:_jmom_dot}), the main change is in the dynamics of the momentum
density, where superfluid velocity drops out, 
\begin{equation}
U\left( \rho _{a}+\rho _{\perp }^{n}\right) v_{x}^{n}-c_{55}\frac{v_{x}^{n}}{%
U}=0.
\end{equation}%
The dispersion equation for the T-waves reads%
\begin{equation}
\left( U^{2}-u_{4\parallel }^{2}\right) \left( U^{2}-\frac{\rho
_{a}u_{55}^{2}}{\rho _{a}+\rho _{\perp }^{n}}\right) =0,
\end{equation}%
where we introduced $\rho _{a}u_{55}^{2}\equiv c_{55}\sim E\psi ^{2}\frac{a}{%
c},$ \ the fourth sound velocity above is 
\begin{equation}
u_{4\parallel }^{2}=u_{1}^{2}\frac{\rho _{\parallel }^{s}}{\rho }%
+u_{2\parallel }^{2}\frac{\rho _{\parallel }^{n}}{\rho },~u_{2\parallel
}^{2}=\frac{Ts^{2}\rho _{\parallel }^{s}}{c_{p}\rho _{\parallel }^{n}}.
\label{eq: u4||,u2||}
\end{equation}
Hence, one gets two degenerate transversal waves with the velocity $%
U_{55}^{T}=u_{55}\left[ \rho _{a}/(\rho _{a}+\rho _{\perp }^{n})\right]
^{1/2}$, \ and two with the fourth sound velocity $U_{a4}^{T}=u_{4\parallel
}.$ Here, we introduced $u_{55}\sim \sqrt{E\psi ^{2}\frac{a}{c}/\rho _{a}}=5$
m/s and the corresponding 'shear' sound velocity near $T_{ca}$ is $%
U_{55}^T\sim 3$ m/s.  This mode is slow enough to be excited in the
mm-size aerogel sample. Note that $U_{55}^T$ exist above critical temperature $T_{ca}$
and may be one of the slow modes that Dmitriev {\em et al.}\cite{dmit20} called 
the 'main' mode that exists in the normal phase and than experiences an 'avoided' 
crossing with the additional mode that emerges at $T_{ca}$, Fig.~\ref{Fig:schema}.

Importantly, the \emph{hybrid fourth sound} velocity 
$U_{a4}^{T}=u_{4\parallel }\rightarrow 0$ vanishes at $T_{ca}$ in contrast
with \emph{hybrid} \emph{longitudinal fourth }sound velocity which remains
finite even at $T=T_{ca}$. One can estimate its behavior near $T_{ca}$ as $%
U_{4a}^{T}\sim u_{1}\sqrt{\frac{\rho _{\parallel }^{s}}{\rho }}\sim u_{1}%
\sqrt{0.3\tau }$ in polar phase. 

For the latter estimate we have used the results for the superfluid density
stemming from the Ginzburg-Landau\ theory of the polar phase. Namely, $\rho
_{\parallel }^{s}/\rho _{\perp }^{s}=3$ in the weak coupling regime, and 
\begin{equation}
	\frac{\rho _{\perp }^{s}}{\rho }=\frac{\tau }{\beta _{12345}},
\end{equation}
 where $\beta
_{12345}=\beta _{1}+...+\beta _{5},$ are the Ginzburg-Landau parameters.
Since the weak coupling approximation is unable to describe the existence of
A-phase and other features of $^{3}$He, one would like to use the \emph{%
strong coupling values} for $\beta $s. For below estimates, we shall use the
value $\rho _{\perp }^{s}/\rho \sim 0.1\tau $ \cite{surfA}.

\subsection{Sound propagating perpendicular to the strands with $\hat {k}%
=(1,0,0)$}

\begin{figure}[ptb]
\centering\includegraphics[width=0.3\textwidth]{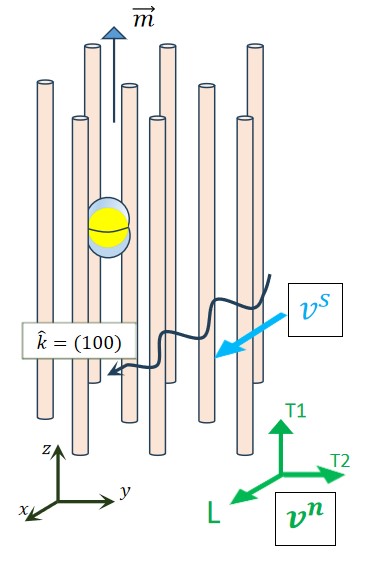} 
\caption{(Color online)
Schematic of sound wave propagating perpendicular to the strands of nematic aerogel filled
with the polar phase  of $^3$He. The superfluid velocity $\textbf{v}^s$
is collinear with the wave  vector $\hat{k}=(1,0,0)$. The normal velocity has one longitudinal and two
transversal possible orientations ($L, T1, T2$),
The modes $T1$ and $T2$ are \emph{not} degenerate: the mode $T1$ involves the normal
density along the strands, $T2$ is perpendicular to the strands and they are different 
reflecting the anisotropy of the superconducting gap.
}\label{Fig:Twave}
\end{figure}

In this case, we again have one longitudinal wave and two transversal waves
in dry aerogel and the additional modes due to the global phase coherence of
superfluid polar phase of $^{3}$He inside aerogel . In variance
with case (A), here the two \emph{transverse waves are not degenerate}.
Indeed, for the velocity of normal motion polarizations go like this: L-wave $%
v^{n}\parallel x=(v_{x}^{n},0,0,)$, T1-wave $v^{n}\parallel
z=(0,0,v_{z}^{n}) $, and T2-wave $v^{n}\parallel y=(0,v_{y}^{n},0,)$, Fig.~%
\ref{Fig:Twave}. Since T1-wave would involve the normal density \emph{%
parallel} to the strands and in the T2-wave one \emph{perpendicular} to
the strands, their velocities would be different. Note that the superfluid
velocity polarization is parallel to the k-vector and thus perpendicular to
the strands, $\boldsymbol{v}^{s}\ \parallel x=(v_{x}^{s},0,0)$ in T-waves. 
To distinguish this case from the previous one for the propagation along the
strands, we will mark the resulting velocities by capital $S.$

\subsubsection{Longitudinal wave propagating perpendicular to the strands}

In this case, the normal and superfluid velocities are collinear and
oscillate perpendicular to the strands $v^{n}\parallel x=(v_{x}^{n},0,0,),$ $%
v_{j}^{s}\parallel x=(v_{x}^{s},0,0)$. The momentum density equation for the
sound velocity $S$\ reads:%
\begin{equation}
\left( \rho _{a}+\rho _{\perp }^{n}\right) Sv^{n}+S\rho _{\perp }^{s}v^{s}-%
\tilde{p}-c_{11}\frac{v^{n}}{S}=0.
\end{equation}%
and the full system gives the dispersion equation for L~-~wave in case $\hat{%
k}=(1,0,0)$%
\begin{align}
\left( S^{2}-u_{1}^{2}\right) \left( S^{2}-u_{2\perp }^{2}\right) +\frac{%
\rho _{a}}{\rho _{\perp }^{n}}\left( S^{2}-u_{11}^{2}\right) \left(
S^{2}-u_{4\perp }^{2}\right) & =0,  \label{eq:B_L} \\
u_{2\perp }^{2}=\frac{Ts^{2}\rho _{\perp }^{s}}{\rho _{\perp }^{n}c_{p}}%
,\quad u_{4\perp }^{2}=u_{1}^{2}\frac{\rho _{\perp }^{s}}{\rho }+u_{2\perp
}^{2}\frac{\rho _{\perp }^{n}}{\rho },&
\end{align}%
where we introduced $\rho _{a}u_{11}^{2}\equiv c_{11}\sim E\psi \frac{a}{c},$
$u_{11}\simeq 20$ m/s. This yields the hybrid second sound $S_{2a}^{L}$ and
fourth sound $S_{4a}^{L}$.  The hybrid fourth sound remains finite and large in
this case as well, $S_{4a}^{L}\sim u_{1}/\sqrt{2}\simeq 200$ m/s. 
Near $T_{ca}$, in full analogy with Eqs.~(\ref{eq:U2a_nearTca}),(\ref{eq:U4a_nearTca}),
\begin{equation}
S^2_{2a} = \tau u_1^2\frac{\rho_a u_{11}^2}{\rho_a u_{11}^2+\rho u_1^2}  ,
\end{equation}%
\begin{equation}
S^2_{4a} = \frac{\rho_a u_{11}^2+\rho u_1^2}{\rho_a+\rho}.
\end{equation}%
 The hybrid second sound $S_{2a}$ starts at $T_{ca}$ and quickly rises before 
hitting the linear sample size cutoff, as illustrated by schematic in Fig.~\ref{Fig:schema}.  

\subsubsection{Transversal wave propagating perpendicular to the strands
(type T1)}

Consider the transversal wave with normal velocity polarized along the
strands, $v^{n}\parallel z=(0,0,v_{z}^{n})$, $v^{s}\parallel
x=(v_{x}^{s},0,0).$ The equation for momentum density takes the form,%
\begin{equation}
-(\rho _{a}+\rho _{\parallel }^{n})Sv_{z}^{n}+c_{55}\frac{v_{z}^{n}}{S}=0.
\end{equation}%
The corresponding dispersion equation is for the fourth sound and the
'shear' mode for bending the nematic aerogel in $\emph{xz-}$plane (we remind
that the $x-$direction is arbitrary in the plane perpendicular to the
strands),%
\begin{equation}
\left( S^{2}-u_{4\perp }^{2}\right) \left( S^{2}-\frac{\rho _{a}u_{55}^{2}}{%
\rho _{a}+\rho _{\parallel }^{n}}\right) =0,
\end{equation}%
where $\rho _{a}u_{55}^{2}=c_{55},$ so that $u_{55}\sim \sqrt{\frac{E\psi
^{2}\frac{a}{c}}{\rho _{a}}}\sim 3$ m/s. \ Thus, the 'shear' sound $%
S_{55}^{T1}=u_{55}\sqrt{\frac{\rho _{a}}{\rho _{a}+\rho _{\parallel }^{n}}}$
perpendicular to the strands is really slow on the order of a few m/s. As
for the fourth sound, $S_{4\perp }^{T1}=u_{4\perp }$, it emerges at $T_{ca},$
\ $S_{4\perp }^{T1}$ $\sim u_{1}\sqrt{\rho _{\perp }^{s}/\rho }\sim u_{1}$\ $%
\sqrt{0.1\tau },$\ and rises very quickly upon slight cooling from $T_{ca}$
until reaching the size cutoff $U_{c}$, {\em similar to the above} $S_{2a}$, Fig.~\ref{Fig:schema}.

\subsubsection{Transversal wave propagating perpendicular to the strands
(type T2{) }}

In the last case that we address, $\hat{k}=(100),$ $\ \hat{v}^{n}\parallel y$%
, \ \ $\hat{v}^{s}\parallel k.$ The momentum density equation reads%
\begin{equation}
\left( \rho _{a}+\rho _{\perp }^{n}\right) Sv_{y}^{n}-c_{66}\frac{v_{y}^{n}}{%
S}=0.
\end{equation}%
and leads to the dispersion equation%
\begin{equation}
\left( S^{2}-u_{4\perp }^{2}\right) \left( S^{2}-\frac{\rho _{a}u_{66}^{2}}{%
\rho _{a}+\rho _{\perp }^{n}}\right) =0,
\end{equation}%
where $\rho _{a}u_{66}^{2}=c_{66}\sim E\psi ^{2}\frac{a}{c},$ 
\begin{equation}
u_{4\perp }^{2}=u_{1}^{2}\frac{\rho _{s\perp }}{\rho }+u_{2\perp }^{2}\frac{%
\rho _{n\perp }}{\rho },\quad u_{2\perp }^{2}=\frac{Ts^{2}\rho _{s\perp }}{%
\rho _{n\perp }c_{p}}  \label{eq:u4_|,u2_|}
\end{equation}%
and the 'shear' sound in nAG has velocity $S_{66}^{T2}=u_{66}\sqrt{\frac{%
\rho _{a}}{\rho _{a}+\rho _{\perp }^{n}}}\sim 3$ m/s.

\begin{figure}[ptb]
\centering\includegraphics[width=0.5\textwidth]{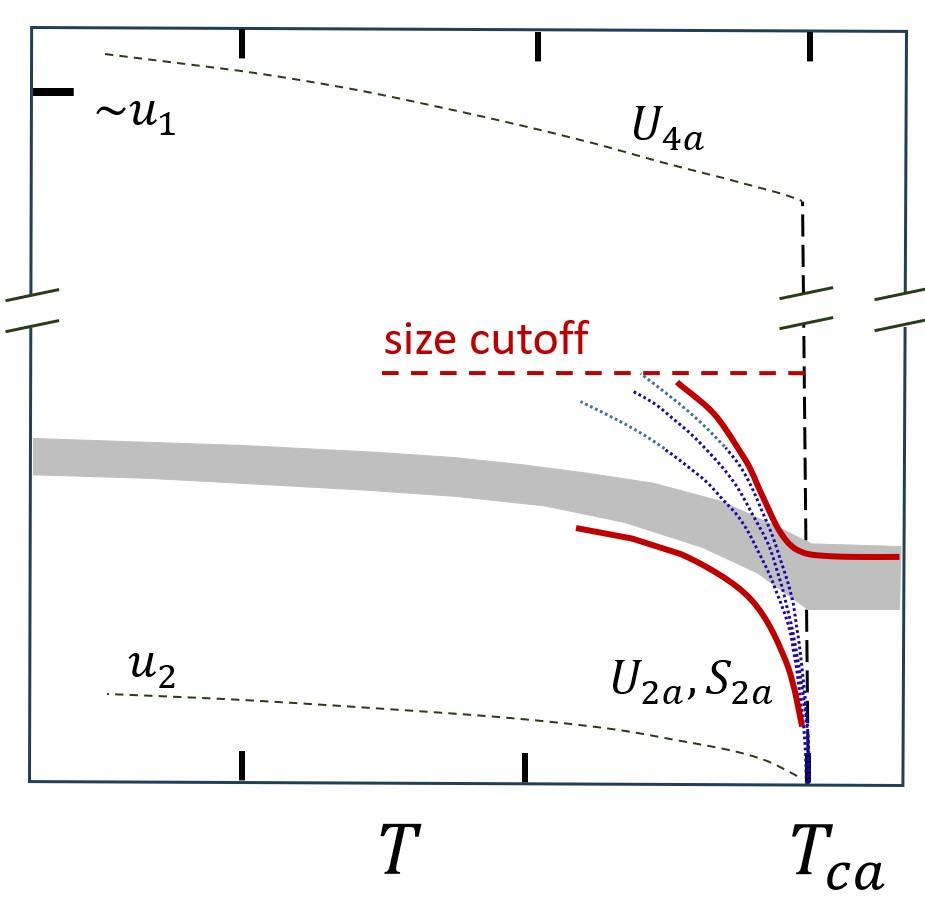} 
\caption{(Color online)
Schematic of the hybrid sound velocities versus temperature in extended nematic aerogel. 
The hybrid second sound $U_{2a}$ and $S_{2a}$ 
emerge at the transition into polar phase with zero velocity. 
Both modes exhibit sharply increasing velocity   
until they hit the size cutoff that grows linearly with the
linear size of aerogel sample (see text.) 
The shear modes due to elastic response of AG skeleton exist above $T_{ca}$ and weakly depend 
on temperature (gray band). 
The observed main VW mode (solid line)  exists above $T_{ca}$ and exhibits an 
avoided crossing with the  additional mode that starts at $T_{ca}$.  The velocity of additional mode 
(upper branch {\em below} $T_{ca}$)
 increases until it approaches the nAG sample size cutoff corresponding to about 1600 Hz \cite{dmit20}. 
}\label{Fig:schema}
\end{figure}

We see again that the transversal modes for sound propagating perpendicular
to the strands of nematic aerogel are very slow, $S_{66}^{T2}$ on the order
of a few m/s. Fourth sound for both $T1$ and $T2$ modes, $S_{4\perp }^{T2}=u_{4\perp
}\sim u_{1}\sqrt{0.1\tau },$  exhibits qualitatively and quantitatively
the same behavior. In both cases the fourth sound vanishes at $T_{ca}$ but
its velocity increases upon cooling very quickly reaching cutoff limit on
the order of $10$ m/s signaled by the plateau in frequency dependence of the
VW resonance\cite{dmit20}, Fig.~\ref{Fig:schema}.

\begin{figure}[ptb]
\centering
\includegraphics[width=0.5\textwidth]{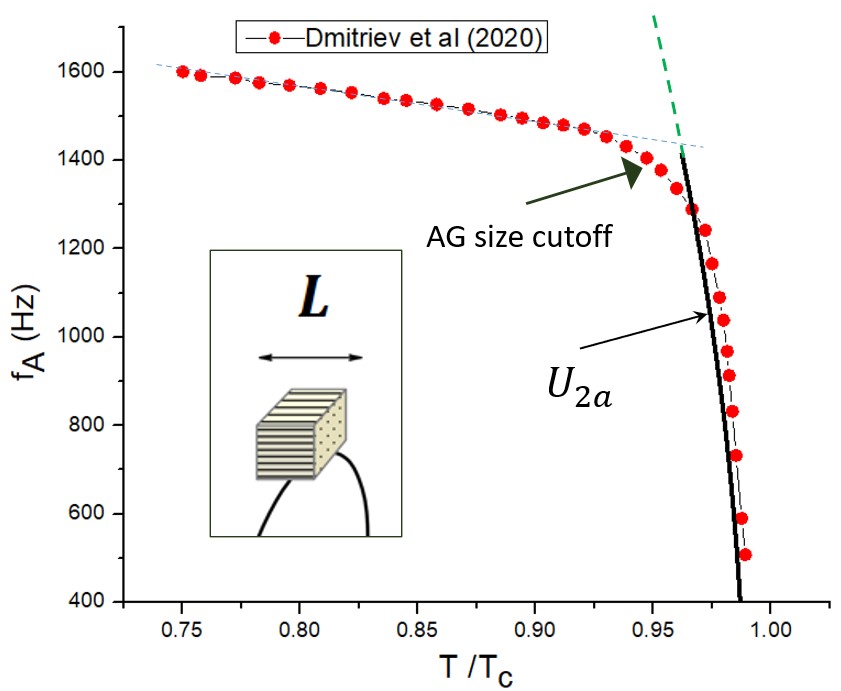} 
\caption{(Color online)
		Comparison of the frequency of hybrid second sound $U_{2a}$ with data \cite{dmit20}.
		The frequency rises sharply from its emergence at $T_{ca}$. It quickly hits a plateau that is
		likely due the velocity of this mode reaching the size cutoff value $U_c \sim 10$ m/s. 
		The size cutoff increases linearly with $L$ (see text).
		Inset shows the aerogel sample with size $L \approx 3$ mm attached to the vibrating wire \cite{dmit20}. 
	}\label{Fig:vdata}
\end{figure}

The broad overview of the above results in presented in Fig.~\ref{Fig:schema}. The main observed mode starts above the transition into the polar phase inside the aerogel sample and continues into the superfluid phase. It is depicted as a gray band since in experiment various modes with different k-vectors  get excited giving the main mode a finite width in addition to dissipation.  A few additional modes emerge at $T_{ca}$ marked by dashed lines. Those interacting with main mode exhibit avoided crossing as shown on schematic right below $T_{ca}$. The lower branch is attributed to the main mode below the transition by Dmitriev {\em at al.} and stays at about 450 Hz\cite{dmit20}. The additional mode (the upper branch) exhibits quick increase in velocity before it approaches a finite sample size cutoff. 

The hybrid fourth sound $U_{4a}$ is peculiar. It does not go to zero at $T_{ca}$ like in bulk $^3$He but retains finite velocity that is on the order of $u_1$, the first sound velocity in $^3$He. It is too fast to get excited in 3mm aerogel sample. Also shown is the velocity of 'ordinary' second sound $u_2$ that is very slow indeed compared to the  AG modes hybridized with the second sound $U_{2a}$ and $S_{2a}$, as discussed above, and is not registered either.  

The hybrid second sound $U_{2a}$ is compared to the data in Fig.~\ref{Fig:vdata}. One observes that the model closely follows the data until  finite size cutoff is hit signaled by a knee in the data for AG sample with $L=3$mm at frequency about 1400 Hz.  Incidentally, the cross over for the plateau varies linearly with $L$ and that may potentially be used in future experiments.

\section{Discussion}

We have shown above that a simple mechanical model of nematic aerogel yields
results for the elastic constants that appear universal since they depend on
few generic parameters: AG porosity (the volume fraction of the solid phase $\psi \approx 5\%$) , 
large aspect ratio of connected network of
strands $c/a \gg1 $, Fig.~\ref{Fig:ucell_w_TEM}, and the Young's modulus of the strands' material. This allows
building a phenomenological elasto-hydrodynamic- model for the emerging
polar phase of $^{3}$He filling up the nematic aerogel and relate dynamics
of sound excitations to the viscous coupling between aerogel strands and the
normal motion of $^3$He and superfluid backflow. 
\begin{table}[t]
	\caption{Sound velocities for $^3$He in nematic aerogel for different
	propagation directions and polarizations. The nAG strands are running along $\hat z$ =(001) direction.}
	\begin{ruledtabular}
		\begin{tabular}[t]{c | c | c | c | c }
			\multicolumn{1}{c|}{$\hat{k}$} & \multicolumn{1}{c|}{$\hat v^s$} & \multicolumn{1}{c|}{$\hat v^{n}$} 
			& \multicolumn{1}{c|}{$U_{2a}^2$} &\multicolumn{1}{c}{$U_{4a}^2$}\\
			\hline \hline
			(001) &(001)  & (001)(L)\footnote{Values close to $T_{ca}$, $\tau=1-T/T_{ca}\ll 1$.
			L stands for the longitudinal, T for the transversely polarized sound waves.
			} 
			& $\tau u_1^2\frac{\rho_a u_{33}^2}{\rho_a u_{33}^2+\rho u_1^2}$ 
			& $\frac{\rho_a u_{33}^2+\rho u_1^2}{\rho_a+\rho}$\\ \cline{3-5}
			&   & (100)(T)  & $ u_{55}^2\frac{\rho_a }{\rho_a+\rho^n_\perp}$ 
			& $u_{4\parallel}^2$\\ \cline{3-5}
			&      & (001)(T)\footnote{The transverse velocities are degenerate for the waves propagating along the strands. } & $-"-$ & $-"-$ \\	
			\hline\hline
			\multicolumn{1}{c|}{$\hat{k}$} & \multicolumn{1}{c|}{$\hat v^s$} & \multicolumn{1}{c|}{$\hat v^n$} & \multicolumn{1}{c|}{$S_{2a}^2$} &\multicolumn{1}{c}{$S_{4a}^2$}\\
			\hline\hline	
			&   &  (100)(L)\footnotemark[1]  &  $\tau u_1^2\frac{\rho_a u_{11}^2}{\rho_a u_{11}^2+\rho u_1^2}$   &$\frac{\rho_a u_{11}^2+\rho u_1^2}{\rho_a+\rho}$ \\ \cline{3-5}
			(100)	&  (100) &  (001)(T1)  & $ u_{55}^2\frac{\rho_a }{\rho_a+\rho^n_\parallel}$   & $u^2_{4\perp}$\\ \cline{3-5}
			&   &  (010)(T2)\footnote{Superfluid motion is perpendicular to the strands for both $T1$ and $T2$ modes 
			resulting in degeneracy of the 4th-like sound velocities. } & $ u_{66}^2\frac{\rho_a }{\rho_a+\rho^n_\perp}$   & $u^2_{4\perp}$\\
		\end{tabular}
	\end{ruledtabular}
\end{table}
We have found that 
(i)
various slow modes with velocities on the order of few meters per second are
excited above and below temperature of transition into the polar phase $T_{ca}$ , 
reflecting the 'softness' of elastic 'shear' (or 'bending') response by nematic aerogel, 
(ii) 
the  second sound hybridized with aerogel vibrations starts with zero velocity at $%
T_{ca} $. 
It then exhibits an avoided crossing with the main mode that persists from the normal phase, and
quickly hits a cutoff imposed by the finite sample size, see discussion of Fig.~\ref{Fig:schema} in the previous Section. This
seems to correlate with observed sharp rise followed by a plateau in
resonant frequency of vibrating wire with lowering $T$ \cite{dmit20,dmitufn24}.
(iii) 
The fourth sound for waves
propagating \emph{perpendicular} \emph{to the strands} , modes $T1$ and $T2$,  also vanish at $T_{ca}$. 
It is worth noting that the hybrid \emph{longitudinal} fourth sound velocity remains finite
	irrespective of direction of propagation  but is too fast to be excited in the
	mm-size aerogel sample. 
 Correspondingly, (iv) there is a size effect cutoff
of hybrid second sound irrespective of directions where it propagates and the
transversal hybrid fourth sound propagating perpendicular to the strands.  All the modes discussed above are classified in the Table.

The current picture offers an alternative view of sound in nematic aerogel
studied previously in similar phenomenological models in Refs.~\cite%
{brand20n22,surfA} yet with large number of free parameters. Here, we have
estimated the elastic constants of the aerogel and found analytical
solutions for all types of sound in the polar $^{3}$He-nAG not available
from prior work. We relate the sharp crossover in observed resonant
frequency to the sample size cutoff so that one has no need in assuming a specific
fine-tuned weak coupling between e.g. $^{3}$He chemical potential oscillations $\tilde{\mu%
}$~(\ref{eq:mu_T_p}) and axial strain $e_{3}$ of the aerogel\cite{surfA}
even though it is allowed by symmetry. More data may shed light onto this interesting interplay.

Obviously, one could not prevent excitation of all modes allowed in aerogel
attached to a vibrating wire and their combination should be responsible
for the observed resonances. In this regard, experiments with vibrations
excited by transducers may be warranted in order to gain more insight into
the above-mentioned size effect and other observed features. As far as 
sharp crossover of resonant frequency with temperature, Fig.~\ref{Fig:vdata}%
, the cutoff sound velocity increases linearly with nAG sample size $L$.
This may be another parameter that one may be able to vary within obvious
experimental constraints.

We acknowledge many enlightening discussions with V.V. Dmitriev, I.A. Fomin,
A.A. Soldatov, E.V. Surovtsev, A.M. Tikhonov, A.M. Troyanovsky, and A.N.
Yudin. V.V. Dmitriev, A.A. Soldatov, and A.N. Yudin are gratefully
acknowledged for providing samples and sharing data.

\section{Data availability} 
All data is included in the main text and is available upon reasonable request.

\end{document}